\documentclass[12pt]{article}
\usepackage{color}
\usepackage{amssymb,latexsym,bm,amsmath,amsthm, amsfonts,multirow,graphicx,mathrsfs}
\usepackage{eufrak,booktabs}
\usepackage{algorithm}
\usepackage{algorithmic}
\usepackage{epsfig}
\usepackage{ulem}
\usepackage{enumitem}
\graphicspath{{figure/}}

\newcommand{\argmin}{\mathop{\rm argmin}}
\newcommand{\argmax}{\mathop{\rm argmax}}

\def\1{{\bm 1}}
\def\0{{\bm 0}}

\def\cov{{\rm cov}}

\def\n{{(n)}}

\renewcommand{\hat}{\widehat}

\newtheorem{thm}{Theorem}

\newtheorem{rmk}[thm]{Remark}

\begin{document}

\title{Robust mislabel logistic regression\\[-1.3ex] without modeling mislabel probabilities}

\author{Hung Hung$^{*1}$, Zhi-Yu Jou$^{1}$, and Su-Yun Huang$^{2}$\\[-1.3ex]
\small $^1$Institute of Epidemiology and Preventive Medicine,
National Taiwan University, Taiwan\\[-1.3ex]
\small $^2$Institute of Statistical Science, Academia Sinica,
Taiwan}

\date{}
\maketitle

\vspace{-0.5cm}
\begin{abstract}
Logistic regression is among the most widely used statistical
methods for linear discriminant analysis. In many applications, we
only observe possibly mislabeled responses. Fitting a conventional
logistic regression can then lead to biased estimation. One common
resolution is to fit a mislabel logistic regression model, which
takes into consideration of mislabeled responses. Another common
method is to adopt a robust $M$-estimation by down-weighting
suspected instances. In this work, we propose a new robust mislabel
logistic regression based on $\gamma$-divergence. Our proposal
possesses two advantageous features: (1) It does not need to model
the mislabel probabilities. (2) The minimum $\gamma$-divergence
estimation leads to a weighted estimating equation without the need
to include any bias correction term, i.e., it is automatically
bias-corrected. These features make the proposed $\gamma$-logistic
regression more robust in model fitting and more intuitive for model
interpretation through a simple weighting scheme. Our method is also
easy to implement, and two types of algorithms are included.
Simulation studies and the Pima data application are presented to
demonstrate the performance of $\gamma$-logistic regression.

\end{abstract}

\clearpage

\section{Introduction}\label{sec.1}

Logistic regression is one of the most widely used statistical
methods for linear discriminant analysis. Let $Y_0$ be a binary
response with $\{0,1\}$ values, and $X$ be the $p$-dimensional
random vector of explanatory variables. Logistic regression assumes
$P(Y_0=1|X=x)$ to satisfy the conditional label probability model
\begin{eqnarray}
\pi(x;\beta)=\frac{\exp(\beta^\top x)}{1+\exp(\beta^\top
x)},\label{model.true}
\end{eqnarray}
where $\beta=(\beta_1,\dots,\beta_p)^\top$, and let $\beta_0$ denote
the true value of $\beta$ in model~(\ref{model.true}). The MLE is
known to be the most efficient estimator for $\beta_0$ when data are
truly generated from model~(\ref{model.true}). However, in some
situations we can only observe a contaminated label $Y$ instead of
the true status $Y_0$. That is, $Y$ is flipped from $Y_0$ according
to the {\it mislabel probabilities}
\begin{eqnarray}
\eta_0(x)=P(Y=1|Y_0=0,X=x)~~ {\rm and}~~ \eta_1(x)=P(Y=0|Y_0=1,X=x).\label{error.prob}
\end{eqnarray}
The success probability of $Y$ no longer follows
model~(\ref{model.true}), but instead has the form
\begin{eqnarray}
P(Y=1|X=x)=\eta_0(x)\, \{1-\pi(x;\beta)\}+\{1-\eta_1(x)\}\,
\pi(x;\beta).\label{model.mis}
\end{eqnarray}
Fitting label contaminated data $\{(Y_i,X_i)\}_{i=1}^n$ to the
uncontaminated model~(\ref{model.true}) will produce a biased
estimate of $\beta_0$. To overcome the problem of mislabeling, some
robustified logistic regression methods are developed based on
(\ref{model.mis}) with different modelings for $\eta_j(x)$'s. Copas
(1988) considered equal and constant mislabel probabilities,
$\eta_0(x)=\eta_1(x)=\eta$, which we call the {\it constant-mislabel
logistic} regression. For any given $\eta$, the estimating equation
of $\beta$ is
$\frac{1}{n}\sum_{i=1}^nw_{\eta,i}(\beta)\{Y_i-\pi_\eta(X_i;\beta)\}X_i=0$
with $\pi_{\eta}(x;\beta) =\eta\,\{1-\pi(x;\beta)\}+(1-\eta)
\,\pi(x;\beta)$ and the weight function
\begin{eqnarray}
w_{\eta,i}(\beta)=\frac{1-2\eta} {\{1-\eta+\eta \exp(-\beta^\top
X_i)\}\{1-\eta+\eta\exp(\beta^\top X_i)\}}. \label{weight.mis}
\end{eqnarray}
Another example is the {\it asymmetric-mislabel logistic} regression
(Wainer, Bradlow and Wang, 2007; Komori {\it et al.}, 2016), which
assumes $\eta_0(x)=\eta$ and $\eta_1(x)=0$, i.e., mislabeling occurs
only in the 0-group. Hayashi (2012) extended the work of
$\eta$-boost (Takenouchi and Eguchi, 2004) to propose a robustified
boosting method for binary classification, which is equivalent to
assuming the following mislabel probabilities
\begin{eqnarray}
\eta_j(x)=\frac{2\xi_j}{(1-\xi_0-\xi_1)\left\{\exp(\frac{1}{2}\beta^\top
x)+\exp(-\frac{1}{2}\beta^\top x)\right\}
+2(\xi_0+\xi_1)},~~~j=0,1\label{eta_mislabel}
\end{eqnarray}
with extra parameters $\xi=(\xi_0,\xi_1)$. We call the corresponding
model the {\it $\xi$-logistic} regression. Note that
(\ref{eta_mislabel}) attains its maximum value at the classification
boundary $\beta^\top x=0$. For any given $\xi$, the estimating
equation of $\beta$ is
$\frac{1}{n}\sum_{i=1}^nw_{\xi,i}(\beta)\{Y_i-\pi_{\xi}(X_i;\beta)\}X_i=0$,
where $\pi_\xi(x;\beta)=\eta_0(x)\,
\{1-\pi(x;\beta)\}+\{1-\eta_1(x)\}\, \pi(x;\beta)$ with $\eta_j(x)$
given in~(\ref{eta_mislabel}), and the weight
\begin{eqnarray}\label{xi_mislabel}
w_{\xi,i}(\beta)=\{1-\eta_0(X_i)-\eta_1(X_i)\}
\nu(X_i;\beta)+\eta_0'(X_i)\{1-\pi(X_i;\beta)\}-\eta_1'(X_i)\pi(X_i;\beta)
\end{eqnarray}
with $\eta_j'(x)=\frac{\partial \eta_j(x)}{\partial(\beta^\top x)}$
and $\nu(x;\beta) = \pi(x;\beta)\{1-\pi(x;\beta)\}$. Robustness of
all the above-mentioned methods come from the underlying weight
functions. For instance, in the constant-mislabel logistic
regression, instances with larger values of $|\beta^\top X_i|$ get
less weight $w_{\eta,i}(\beta)$ in the estimating equation.

The MLE for the above-mentioned robust logistic regression models,
where mislabel probabilities $\eta_j(x)$'s are assumed to take a
certain parametric form, is known to be sensitive to the
misspecification of $\eta_j(x)$'s. Modeling mislabel probabilities,
however, may not be straightforward. In applications, often we are
mainly interested in is the true success probability $P(Y_0=1|X)$
instead of the nuisance parameters $\eta_j(x)$'s. There seems to be
less necessary to build models for $\eta_j(x)$'s. The aim of this
paper is to develop a robust mislabel logistic inference procedure
that avoids modeling $\eta_j(x)$'s. The main idea is to replace the
minimum Kullback-Leibler (KL) divergence estimation, which
corresponds to the MLE, with the minimum $\gamma$-divergence
estimation, which we call {\it $\gamma$-logistic} regression.

The paper is organized as follows. In Section~2, we review
$\gamma$-divergence and use it to propose our robust
$\gamma$-logistic regression, while its asymptotic properties and
comparisons with existing methods are discussed in Section~3.
Simulation studies and the Pima data analysis are placed in
Sections~4-5. The paper ends with a discussion in Section~6.

\section{Method: $\gamma$-Logistic Regression}\label{sec.review}

\subsection{The minimum $\gamma$-divergence estimation and its robustness}\label{sec.div}

Let $g$ be the data generating distribution and $f_\theta$ be the
model distribution indexed by the parameter $\theta$, and let
$\theta_0$ denote the true parameter value of interest. The
$\gamma$-divergence between $g$ and $f_\theta$ is defined to be
\begin{eqnarray}
D_\gamma(g,f_\theta)= \frac1{\gamma(\gamma+1)}
\left\{\|g\|_{\gamma+1} -\int
\Big(\frac{f_\theta}{\|f_\theta\|_{\gamma+1}}\Big)^\gamma g\right\},
\label{div.gamma}
\end{eqnarray}
where $\|f_\theta\|_{\gamma+1}=(\int
f_\theta^{\gamma+1})^{\frac{1}{\gamma+1}}$.  This divergence is
introduced in Jones {\it et al.} (2001) with the name {\it density
power divergence of type-zero}. The name {\it $\gamma$-divergence}
is later introduced in Fujisawa and Eguchi (2008). In the limiting
case, $\lim_{\gamma\to
0}D_\gamma(g,f_\theta)=\int\ln(\frac{g}{f_\theta})g$, which is the
KL-divergence. The estimation criterion of minimum
$\gamma$-divergence estimates $\theta_0$ by
\begin{eqnarray}
\argmin_\theta D_\gamma(g,f_\theta) =\argmax_\theta \int
\Big(\frac{f_\theta}{\|f_\theta\|_{\gamma+1}}\Big)^\gamma
g.\label{gamma.estimation}
\end{eqnarray}
When $g$ belongs to the parametric class
$\{f_{\theta}:\theta\in\Theta\}$ with the parameter value
$\theta_0$, the problem~(\ref{gamma.estimation}) is optimized at
$\theta=\theta_0$. It ensures the consistency of the minimum
$\gamma$-divergence estimation. In the presence of contamination,
however, $g=cf_{\theta_0}+(1-c)h$ which is a mixture of the target
distribution $f_{\theta_0}$ and certain contamination distribution
$h$, where $1-c$ denotes the contamination proportion. With some
calculations, it leads to
\begin{eqnarray}
D_\gamma(g,f_\theta)=\left\{ c\, D_\gamma(f_{\theta_0},
f_\theta)+\frac{B_\gamma(c,h;\theta)}{\gamma(\gamma+1)}\right\}
 + \frac{\|cf_{\theta_0}+(1-c)h\|_{\gamma+1}-
c\|f_{\theta_0}\|_{\gamma+1}}{\gamma(\gamma+1)}\label{div.gamma.contaminated}
\end{eqnarray}
with $B_\gamma(c,h;\theta)=(1-c)\int
(\frac{f_{\theta}}{\|f_\theta\|_{\gamma+1}})^\gamma h$. Ignoring
terms not involving $\theta$, minimizing
(\ref{div.gamma.contaminated}) over $\theta$ is equivalent to
minimizing
\begin{equation}\label{div.gamma.approx}
c\, D_\gamma(f_{\theta_0},
f_\theta)+\frac{B_\gamma(c,h;\theta)}{\gamma(\gamma+1)} \approx c\,
D_\gamma(f_{\theta_0},f_\theta),
\end{equation}
where the approximation holds provided that, for some $\gamma$, the
bias $B_\gamma(c,h;\theta)$ is negligibly small for $\theta$ in a
neighborhood of $\theta_0$. The right hand side of
(\ref{div.gamma.approx}) is minimized at $\theta=\theta_0$. That is,
the minimization process is less affected by the mixing proportion
$c$ and the contamination $h$ and, hence, we can estimate $\theta_0$
well with negligibly small bias. See Fujisawa and Eguchi (2008) and
Kanamori and Fujisawa (2015) for further discussions.

\subsection{$\gamma$-Logistic Regression}\label{sec.gamma.logistic}

The robust $\gamma$-divergence can be used to infer
model~(\ref{model.true}) when the data are actually generated
from~(\ref{model.mis}). The reasons are discussed below.

\begin{thm}\label{thm.mixture}
The distribution of contaminated $Y$ in~(\ref{model.mis}) can be
expressed as a mixture of the target distribution $P(Y_0=y|X=x)$ and
the mislabel-induced distribution $h(y|x)$,
\begin{eqnarray*}
P(Y=y|X=x) = c(x)\, P(Y_0=y|X=x)+\{1-c(x)\}\, h(y|x),
\end{eqnarray*}
where $h(y|x)=\left\{\frac{\eta_0(x)}{\eta_0(x)+\eta_1(x)}\right\}^y
\left\{\frac{\eta_1(x)}{\eta_0(x)+\eta_1(x)}\right\}^{1-y}$ and\,
$1-c(x)=\eta_0(x)+\eta_1(x)$ is the conditional contamination
proportion given $X=x$.
\end{thm}

Theorem~\ref{thm.mixture} sheds some light on the possibility of
inferring the true success probability $P(Y_0=y|X)$ from the
contaminated data $(Y,X)$, since $\gamma$-divergence is able to
ignore the influence from $h(y|x)$ as revealed
in~(\ref{div.gamma.approx}). Specifically, our robust
$\gamma$-logistic adopts the conventional logistic regression model
\begin{eqnarray}
f(y|x;\beta)=\{\pi(x;\beta)\}^y\{1-\pi(x;\beta)\}^{1-y}\label{model.pmf}
\end{eqnarray}
for $Y_0$, while the observed $Y$ is generated from
(\ref{model.mis}), or equivalently, from the mixture
\begin{eqnarray}
g(y|x)=c(x)\, f(y|x;\beta_0)+\{1-c(x)\}\, h(y|x),\label{g}
\end{eqnarray}
where $c(x)$ and $h(y|x)$ are defined in Theorem~\ref{thm.mixture}.
By substituting the model distribution $f(y|x;\beta)$ and the data
distribution $g(y|x)$ into (\ref{gamma.estimation}) and taking
expectation with respect to $X$, $\gamma$-logistic estimates
$\beta_0$ by
\begin{eqnarray}
\argmin_\beta
E_X\Big[D_\gamma\Big\{g(\cdot|X),f(\cdot|X;\beta)\Big\}\Big]
=\argmax_\beta E_{X,Y}\Big\{\Big(\frac{f(Y|X;\beta)}
  {\|f(\cdot|X;\beta)\|_{\gamma+1}}\Big)^\gamma\Big\},\label{gamma_div.reg}
\end{eqnarray}
where $\left\|f(\cdot|x;\beta)\right\|_{\gamma+1}
=[\{\pi(x;\beta)\}^{\gamma+1}+
\{1-\pi(x;\beta)\}^{\gamma+1}]^{1/(\gamma+1)}$, and $E_X$ and
$E_{X,Y}$ denote the expectation with respect to $X$ and $(X,Y)$,
respectively. Recall that the validity of minimum
$\gamma$-divergence estimation relies on the
approximation~(\ref{div.gamma.approx}), where the bias term
$B_\gamma(c,h;\beta)$ plays a key role. From the expressions of
$f(y|x;\beta)$ in (\ref{model.pmf}) and $(c(x), h(y|x))$ in
Theorem~\ref{thm.mixture}, we derive in Supplementary Materials that
\begin{eqnarray}
B_\gamma\{c(x),h(\cdot|x);\beta\}&=&\eta_0(x)\Big\{\pi(x;(\gamma+1)\beta)\Big\}^{\frac{\gamma}{\gamma+1}}+
\eta_1(x)\Big\{1-\pi(x;(\gamma+1)\beta)\Big\}^{\frac{\gamma}{\gamma+1}}\nonumber\\
&\to& \eta_0(x)I(\beta^\top x>0)+ \eta_1(x)I(\beta^\top x\le0)~~ \mbox{as
$\gamma\to\infty$},\label{div.gamma.B}
\end{eqnarray}
where $I(\cdot)$ is an indicator function. It implies that the
robustness of $\gamma$-logistic can be ensured for a large~$\gamma$,
provided that $E_X\{\eta_0(X)I(\beta^\top X>0)\}$ and
$E_X\{\eta_1(X)I(\beta^\top X\le 0)\}$ at $\beta\approx\beta_0$ are
negligible, under which $E_X[B_\gamma\{c(X),h(\cdot|X);\beta\}]$ can
only have limited influence on (\ref{gamma_div.reg}). See
Remarks~\ref{rmk.robsutness.gamma}-\ref{rmk.mis_model} for further
discussions about the robustness of $\gamma$-logistic.

In the sample level, the robust estimator $\widehat\beta_\gamma$ is
obtained via the empirical version of~(\ref{gamma_div.reg}),
\begin{eqnarray}
\frac{1}{n}\sum_{i=1}^n\Big(\frac{f(Y_i|X_i;\beta)}
{\|f(\cdot|X_i;\beta)\|_{\gamma+1}}\Big)^\gamma =
\frac{1}{n}\sum_{i=1}^n\left(\frac{\exp\{Y_i\,(\gamma+1)\beta^\top
X_i\}} {1+\exp\{(\gamma+1)\beta^\top
X_i\}}\right)^{\frac{\gamma}{\gamma+1}}.\label{gamma_div.reg.sample}
\end{eqnarray}
Direct differentiation of (\ref{gamma_div.reg.sample}) leads to the
estimating equation $S_\gamma(\widehat\beta_\gamma)=0$, where
\begin{eqnarray}
S_\gamma(\beta)&=&\frac{1}{n}\sum_{i=1}^n w_{\gamma,i}(\beta)
\Big\{Y_i-\pi(X_i;(\gamma+1)\beta)\Big\} X_i
\label{estimating_eq.gamma}
\end{eqnarray}
with the weight function
\begin{eqnarray}
w_{\gamma,i}(\beta)&=&\left(\frac{\exp\{Y_i\,(\gamma+1)\beta^\top
X_i\}} {1+\exp\{(\gamma+1)\beta^\top
X_i\}}\right)^{\frac{\gamma}{\gamma+1}}.\label{wieght.gamma}
\end{eqnarray}
From (\ref{estimating_eq.gamma})-(\ref{wieght.gamma}), the
robustness of $\widehat\beta_\gamma$ is clear, as
$w_{\gamma,i}(\beta)$ down-weights instances with non-matched
$(Y_i,\beta^\top X_i)$. Note that the robustness of
$\gamma$-logistic is controlled by the value of~$\gamma$. When
$\gamma=0$, the estimating equation reduces to the non-robust
estimating equation, $\frac{1}{n}\sum_{i=1}^n
\{Y_i-\pi(X_i;\beta)\}X_i=0$, for the conventional logistic
regression. On the other hand, a large $\gamma$ corresponds to a
robust estimate of $\beta_0$, but at the cost of being less
efficient than MLE. See Remark~\ref{rmk.select_gamma} for a
selection method of $\gamma$. See also Supplementary Materials for
two types of algorithms for implementing $\gamma$-logistic
regression.

Besides the parameter estimation, another important issue is to
identify mislabeled subjects. Kanamori and Fujisawa (2015) developed
a method to estimate the expected mixing proportion, $c = E_X\{
c(X)\}$, based on the density power divergence. With this estimated
$c$, they proposed to identify $100(1-c)\%$ subjects with the
smallest estimated values of $f(Y_i|X_i;\beta_0)$ as outliers. On
the other hand, the weight $w_{\gamma,i}(\widehat\beta_\gamma)$ from
$\gamma$-logistic can provide a measure of label confidence. It
motivates us to identify mislabeled subjects by searching for
instances with small values of $w_{\gamma,i}(\widehat\beta_\gamma)$.
To have an objective evaluation criterion, we obtain the p-values of
$w_{\gamma,i}(\widehat\beta_\gamma)$'s by parametric bootstrap. Let
$w_{\gamma,i}^{(b)}$ be the $b$-th bootstrapped version of
$w_{\gamma,i}(\widehat\beta_\gamma)$ by the null data
$\{(\widehat{Y}_i^{(b)},X_i)\}_{i=1}^n$, where $\widehat{Y}_i^{(b)}$
is generated from model~(\ref{model.true}) given $X=X_i$ and
$\beta=\widehat\beta_\gamma$. The p-value of
$w_{\gamma,i}(\widehat\beta_\gamma)$ is
\begin{eqnarray}
PV_i=\frac{1}{b'}\sum_{b=1}^{b'}I\left\{w_{\gamma,i}^{(b)}\le
w_{\gamma,i}(\widehat\beta_\gamma)\right\}\label{PV}
\end{eqnarray}
for a large $b'$. Instances, e.g., $\{i:PV_i<0.01\}$, can be
identified for further examination.

We close this section by giving a few remarks on the robustness of
$\gamma$-logistic, the confounding issue of the model
misspecification and mislabeling, and the selection of $\gamma$
value.

\begin{rmk}[robustness]\label{rmk.robsutness.gamma}
For symmetric mislabeling $\eta_0(x)=\eta_1(x)$,
equation~(\ref{div.gamma.B}) becomes
$\lim_{\gamma\to\infty}B_\gamma\{c(x),h(\cdot|x);\beta\}=\eta_0(x)=\eta_1(x)$,
which does not involve the parameter $\beta$, i.e., the bias term
$B_\gamma\{c(x),h(\cdot|x);\beta\}$ plays no role in parameter
estimation as $\gamma\to\infty$. In other words, $\gamma$-logistic
with a large $\gamma$ produces a consistent estimate of $\beta_0$
regardless of the functional forms of $\eta_0(x)$ and $\eta_1(x)$ as
long as $\eta_0(x)=\eta_1(x)$.
\end{rmk}

\begin{rmk}[confounding]\label{rmk.mis_model}
In all previous discussions, we assume the model is correctly
specified, i.e, $P(Y_0=1|X=x)=\pi(x;\beta_0)$ for some $\beta_0$.
When the model is misspecified, there is no so-called true
$\beta_0$, and the target parameter is criterion-dependent. With
$\gamma$-divergence, the target parameter is
$\beta_\gamma^*=\argmin_\beta E_X
[D_\gamma\{f_{Y_0|X}(\cdot|X),f(\cdot|X;\beta)\}]$ with
$f_{Y_0|X}(y|x)=P(Y_0=y|X=x)$. Similar to the derivation of
(\ref{div.gamma.contaminated}) with $g(y|x)=c(x)
f_{Y_0|X}(y|x)+\{1-c(x)\}h(y|x)$, we have (up to terms without
involving $\beta$)
\begin{eqnarray*}
D_\gamma\{g(\cdot|x),f(\cdot|x;\beta)\}\propto c(x)\,
D_\gamma\{f_{Y_0|X}(\cdot|x),
f(\cdot|x;\beta)\}+\frac{B_\gamma\{c(x),h(\cdot|x);\beta\}}{\gamma(\gamma+1)}
\end{eqnarray*}
with the bias term $B_\gamma\{c(x),h(\cdot|x);\beta\}$ being defined
in (\ref{div.gamma.B}). Note that
$B_\gamma\{c(x),h(\cdot|x);\beta\}$ does not involve
$f_{Y_0|X}(y|x)$, and the robustness of $\gamma$-logistic in
estimating $\beta_\gamma^*$ is still valid for a large $\gamma$,
provided that $E_X\{\eta_0(X)I(\beta^\top X>0)\}$ and
$E_X\{\eta_1(X)I(\beta^\top X\le 0)\}$ at $\beta\approx
\beta_\gamma^*$ are small as discussed in texts below
(\ref{div.gamma.B}). Moreover, by Remark~\ref{rmk.robsutness.gamma},
the robustness of $\gamma$-logistic is unaffected by the functional
forms of $\eta_0(x)$ and $\eta_1(x)$ when $\eta_0(x)=\eta_1(x)$.
\end{rmk}

\begin{rmk}[selection of $\gamma$]\label{rmk.select_gamma}
One can use the idea of Mollah, Eguchi and Minami (2007) to select
$\gamma$ by $\argmax_\gamma \,\frac{1}{n}\sum_{i=1}^n
w_{\gamma_0,i}(\widehat\beta_\gamma)$ from
(\ref{gamma_div.reg.sample}), where $\gamma_0$ is a predetermined
reference value, e.g., $\gamma_0=0.1$. Note that
(\ref{gamma_div.reg.sample}) can be affected by mislabeled $Y_i$.
Thus, alternatively we replace the weight by its conditional
expectation, i.e., $E[w_{\gamma_0,i}(\beta_0)|X_i]
=\|f(\cdot|X_i;\beta_0)\|_{\gamma_0+1}$, and propose to select
$\gamma$ by $\argmax_\gamma\,
\frac{1}{n}\sum_{i=1}^n\|f(\cdot|X_i;\widehat\beta_\gamma)\|_{\gamma_0+1}$.
\end{rmk}

\section{Characteristics of $\gamma$-Logistic Regression}\label{sec.comparison}

\subsection{Influence function and asymptotic properties of
$\widehat\beta_\gamma$}\label{sec.asymptotic}

Since $\widehat\beta_\gamma$ is an $M$-estimator, the influence
function ${\sf IF}_{\widehat\beta_\gamma}(X_i,Y_i)$ of
$\widehat\beta_\gamma$ evaluated at $(Y_i,X_i)$ and $\beta=\beta_0$
is the negative Hessian inverse times the $i$-th element of the
score function:
\begin{eqnarray}
{\sf IF}_{\widehat\beta_\gamma}(X_i,Y_i)=
w_{\gamma,i}(\beta_0)\,\Big\{Y_i-\pi(X_i;(\gamma+1)\beta_0)\Big\}\,
H_\gamma^{-1}X_i, \label{IF.gamma}
\end{eqnarray}
where $H_\gamma=E[-\frac{\partial}{\partial\beta}S_\gamma(\beta)
  |_{\beta=\beta_0}]=E[w_{\gamma,i}(\beta_0)\,\nu(X_i;(\gamma+1)\beta_0)\,
X_iX_i^\top]+\Delta_\gamma$ with
\begin{eqnarray*}
\Delta_\gamma=\gamma \, E\Big[w_{\gamma,i}(\beta_0)\,\Big[
\nu(X_i;(\gamma+1)\beta_0)
-\Big\{Y_i-\pi(X_i;(\gamma+1)\beta_0)\Big\}^2\Big]\,
X_iX_i^\top\Big]
\end{eqnarray*}
and $\nu(x;\beta) = \pi(x;\beta)\{1-\pi(x;\beta)\}$. Direct
calculation gives $\Delta_\gamma=0$, and $H_\gamma$ reduces to
\begin{eqnarray}
H_\gamma=E\Big[\|f(\cdot|X_i;\beta_0)\|_{\gamma+1}\,
\nu(X_i;(\gamma+1)\beta_0)\, X_iX_i^\top\Big]. \label{H.w2}
\end{eqnarray}
The robustness of $\gamma$-logistic can be seen from ${\sf
IF}_{\widehat\beta_\gamma}(X_i,Y_i)$, where a large difference
$\{Y_i-\pi(X_i;(\gamma+1)\beta_0)\}$ (which occurs when $Y_i$ is
mislabeled) will accompany with a small value of
$w_{\gamma,i}(\beta_0)$, so that the influence of mislabeling is
mitigated. As to the case of conventional logistic regression, which
corresponds to ${\sf IF}_{\widehat\beta_\gamma}(X_i,Y_i)$ with
$\gamma=0$, we have $w_{\gamma,i}(\beta_0)=1$ and there is no chance
to achieve robustness when $Y_i$ is mislabeled.

The asymptotic normality of $\gamma$-logistic is established below.
\begin{thm}\label{thm.gamma}
Assume the validity of model~(\ref{model.true}) and $E\|{\sf
IF}_{\widehat\beta_\gamma}(X,Y)\|^2<\infty$. As $n\to\infty$, we
have the weak convergence
$\sqrt{n}(\widehat\beta_\gamma-\beta_0)\stackrel{d}{\rightarrow}N(0,\Sigma_\gamma)$,
where $\Sigma_\gamma=H_\gamma^{-1}U_\gamma H_\gamma^{-1}$,
$H_\gamma$ is defined in (\ref{H.w2}), and
$U_\gamma=E[w_{\gamma,i}^2(\beta_0)\{Y_i-\pi(X_i;(\gamma+1)\beta_0)\}^2X_iX_i^\top]$.
\end{thm}

\noindent The asymptotic covariance matrix $\Sigma_\gamma$ can be
estimated by the sandwich-type estimator
\begin{eqnarray}\label{cov.est}
\widehat\Sigma_\gamma=\Big\{\widehat
H_\gamma(\widehat\beta_\gamma)\Big\}^{-1}\,\widehat
U_\gamma(\widehat\beta_\gamma)\, \Big\{\widehat
H_\gamma(\widehat\beta_\gamma)\Big\}^{-1},
\end{eqnarray}
where
\begin{eqnarray*}
{\widehat
U}_\gamma(\beta)&=&\frac{1}{n}\sum_{i=1}^nw_{\gamma,i}^2(\beta)
\{Y_i-
\pi(X_i;(\gamma+1)\beta)\}^2 X_iX_i^\top\\
\widehat H_\gamma(\beta)
&=&\frac{1}{n}\sum_{i=1}^n\|f(\cdot|X_i;\beta)\|_{\gamma+1}
\nu(X_i;(\gamma+1)\beta)\, X_iX_i^\top
+\widehat\Delta_\gamma(\beta)\\
\widehat\Delta_\gamma(\beta)&=&
\frac{\gamma}{n}\sum_{i=1}^nw_{\gamma,i}(\beta)\,\big[
\nu(X_i;(\gamma+1)\beta)
-\big\{Y_i-\pi(X_i;(\gamma+1)\beta)\big\}^2\big]\, X_iX_i^\top.
\end{eqnarray*}
Note that we still include
$\widehat\Delta_\gamma(\widehat\beta_\gamma)$ in $\widehat
H_\gamma(\widehat\beta_\gamma)$ to estimate $\Delta_\gamma=0$, since
its effect cannot be ignored under finite samples. Subsequent
inference about $\beta_0$ can be based on $(\widehat\beta_\gamma,
\widehat\Sigma_\gamma)$.

\subsection{Comparison with model-based mislabel logistic regression}

\begin{figure}[!ht]
\hspace{-0.3in}
\begin{center}
\includegraphics[width=3in]{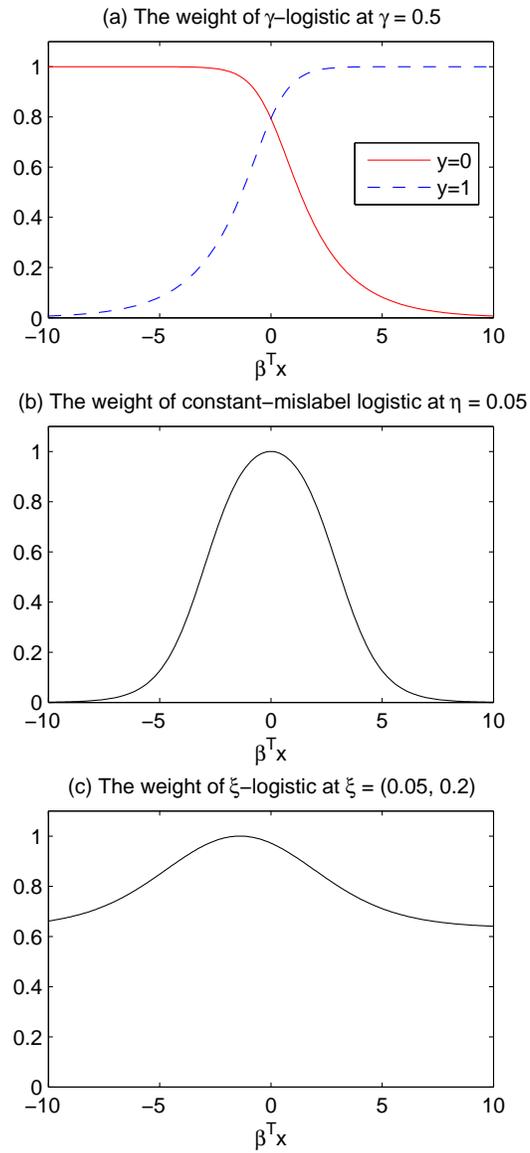}
\end{center}
\caption{The weight functions (scaled to have a maximum value 1) of
(a)~$\gamma$-logistic with $\gamma=0.5$, (b)~constant-mislabel
logistic with $\eta=0.05$, and (c)~$\xi$-logistic with
$\xi=(0.05,0.2)$.}\label{fig.weight}
\end{figure}

A major difference between $\gamma$-logistic and the model-based
mislabel logistic, e.g., constant-mislabel logistic and
$\xi$-logistic, is the weight functions (see
Figure~\ref{fig.weight}). The weights $w_{\eta,i}(\beta)$ and
$w_{\xi,i}(\beta)$ depend on $\beta^\top X_i$ only, which always
down-weight samples with large $|\beta^\top X_i|$ values. Among
these instances with large $|\beta^\top X_i|$ values, some are
correctly-labeled. On the other hand, the weight
$w_{\gamma,i}(\beta)$ of $\gamma$-logistic depends on both
$(Y_i,X_i)$, and it only down-weights instances with non-matched
$(Y_i,\beta^\top X_i)$. $\gamma$-logistic is able to weigh data
instances in a more correct way, and thus can be expected to perform
better than model-based mislabel logistic regressions under severe
contamination. Another advantage is that the validity of
$\gamma$-logistic mainly relies on putting less weight on instances
having non-matched $(Y_i,\beta^\top X_i)$, and does not rely on any
modeling of the mislabel probabilities $\eta_j(x)$'s. As for
model-based mislabel logistic regressions, they incorporate the
mislabel probabilities into model~(\ref{model.mis}), which requires
a further modeling for the nuisance parameters $\eta_j(x)$'s. The
form of $\eta_j(x)$, however, is rarely known in practice, and the
performance of model-based mislabel logistic can be questionable
when complicated mislabel probabilities are present.

\subsection{Comparison with robust mislabel logistic regression
using density power divergence}\label{sec.comparison.alpha}

Ghosh and Basu (2016) proposed a robust GLM by the minimum density
power divergence estimation. For any $\alpha>0$, the density power
divergence between $g$ and $f_\theta$ is
\begin{eqnarray}
D_\alpha(g,f_\theta)=\alpha\,\int
f_\theta^{\alpha+1}-(\alpha+1)\,\int gf_\theta^\alpha+\int
g^{\alpha+1}.\label{div.alpha}
\end{eqnarray}
The estimating equation by replacing $D_\gamma$
in~(\ref{gamma_div.reg}) with $D_\alpha$ becomes
$S_\alpha(\widehat\beta_\alpha)=0$, where
\begin{eqnarray}
S_\alpha(\beta)=\frac{1}{n}\sum_{i=1}^n\left[
w_{\alpha,i}(\beta)\Big\{Y_i-\pi(X_i;\beta)\Big\}-
b_\alpha(X_i;\beta) \right] X_i \label{estimating_eq.alpha}
\end{eqnarray}
with the weight $w_{\alpha,i}(\beta) =
\left\{\frac{\exp(Y_i\,\beta^\top X_i)}{1+\exp(\beta^\top
X_i)}\right\}^\alpha$ and $b_\alpha (x;\beta)=\frac{\exp(\beta^\top
x) \{\exp(\alpha\beta^\top x)-1\}} {\{1+\exp(\beta^\top
x)\}^{2+\alpha}}$ being the bias correction term. The following
result is established by Ghosh and Basu (2016).

\begin{thm}[Ghosh and Basu, 2016] Under model~(\ref{model.true}), the influence function of
$\widehat\beta_\alpha$ evaluated at $(Y_i,X_i)$ and $\beta=\beta_0$
is ${\sf IF}_{\widehat\beta_\alpha}(X_i,Y_i)=[
w_{\alpha,i}(\beta_0)\{Y_i-\pi(X_i;\beta_0)\}- b_\alpha(X_i;\beta_0
)]H_{\alpha}^{-1}X_i,$ where $ H_\alpha=E[\xi_\alpha(X_i;\beta_0)\,
\nu(X_i;\beta_0) \, X_iX_i^\top ]$ with
$\xi_\alpha(x;\beta)=\frac{\exp(\alpha\beta^\top x)+\exp(\beta^\top
x)} {\{1+\exp(\beta^\top x)\}^{1+\alpha}}$. Moreover,
$\sqrt{n}(\widehat\beta_\alpha-\beta_0)\stackrel{d}{\rightarrow}N(0,\Sigma_\alpha)$
with $\Sigma_\alpha=H_\alpha^{-1}U_\alpha H_\alpha^{-1}$ and $
U_\alpha=E[\xi^2_\alpha(X_i;\beta_0)\, \nu(X_i;\beta_0) \,
X_iX_i^\top ]$.
\end{thm}

For simplicity in notation, we use the term {\it $\alpha$-logistic}
to denote the Ghosh-Basu logistic regression, since the density
power divergence $D_\alpha$ is indexed by $\alpha$. Although both
$\gamma$-logistic and $\alpha$-logistic are derived from the minimum
divergence estimation, they have different behaviors in estimating
$\beta_0$. First, the robustness of both methods comes from the
weight functions $w_{\gamma,i}(\beta)$ and $w_{\alpha,i}(\beta)$,
and they are connected via
$\{w_{\gamma,i}(\beta)\}^{\gamma+1}=w_{\alpha,i}((\gamma+1)\beta)$
when $\gamma=\alpha$. It indicates that the two methods share the
same spirit of robustness. However, the resulting estimating
equations are quite different in the bias correction scheme. In
particular, $\gamma$-logistic corrects the bias by using the
expanded parameter $(\gamma+1)\beta$ in~(\ref{estimating_eq.gamma}),
while $\alpha$-logistic subtracts a bias correction term
$b_\alpha(x;\beta)$ in~(\ref{estimating_eq.alpha}). A consequence is
that $S_\gamma(\beta)$ of $\gamma$-logistic consists of a weighted
sum expression with the weight $w_{\gamma,i}(\beta)$, which directly
reflects the contribution of the $i$-th instance to the estimator
$\widehat\beta_\gamma$, while this is not the case for
$S_\alpha(\beta)$ of $\alpha$-logistic. Another difference is the
ability of robustness. As shown in~(\ref{div.gamma.approx}),
$\gamma$-divergence is able to ignore the influence of mislabeling,
and we can expect a strong robustness property for
$\gamma$-logistic. However, this is not the case for the density
power divergence $D_\alpha$. This can be seen from the fact that,
when $g=cf_{\theta_0}+(1-c)h$, we have
\begin{eqnarray}
D_\alpha(g,f_\theta) ~\propto~ D_\alpha(cf_{\theta_0},f_\theta)
-(1-c)\,\int f_{\theta}^\alpha ~\approx~
D_\alpha(cf_{\theta_0},f_\theta),\label{div.alpha.approx}
\end{eqnarray}
where the approximation holds provided that $(1-c)\int
f_{\theta}^\alpha h$ is small enough (Kanamori and Fujisawa, 2015).
Unlike $D_\gamma(g,f_\theta)$ in (\ref{div.gamma.approx}), where the
mixing proportion $c$ appears outside
$D_\gamma(f_{\theta_0},f_\theta)$, here the mixing proportion $c$
appears inside $D_\alpha(cf_{\theta_0},f_\theta)$. This effect leads
to less robustness of $\alpha$-logistic compared with
$\gamma$-logistic.

\begin{figure}[!ht]
\hspace{-0.9in}
\includegraphics[height=6.5in,width=7.8in]{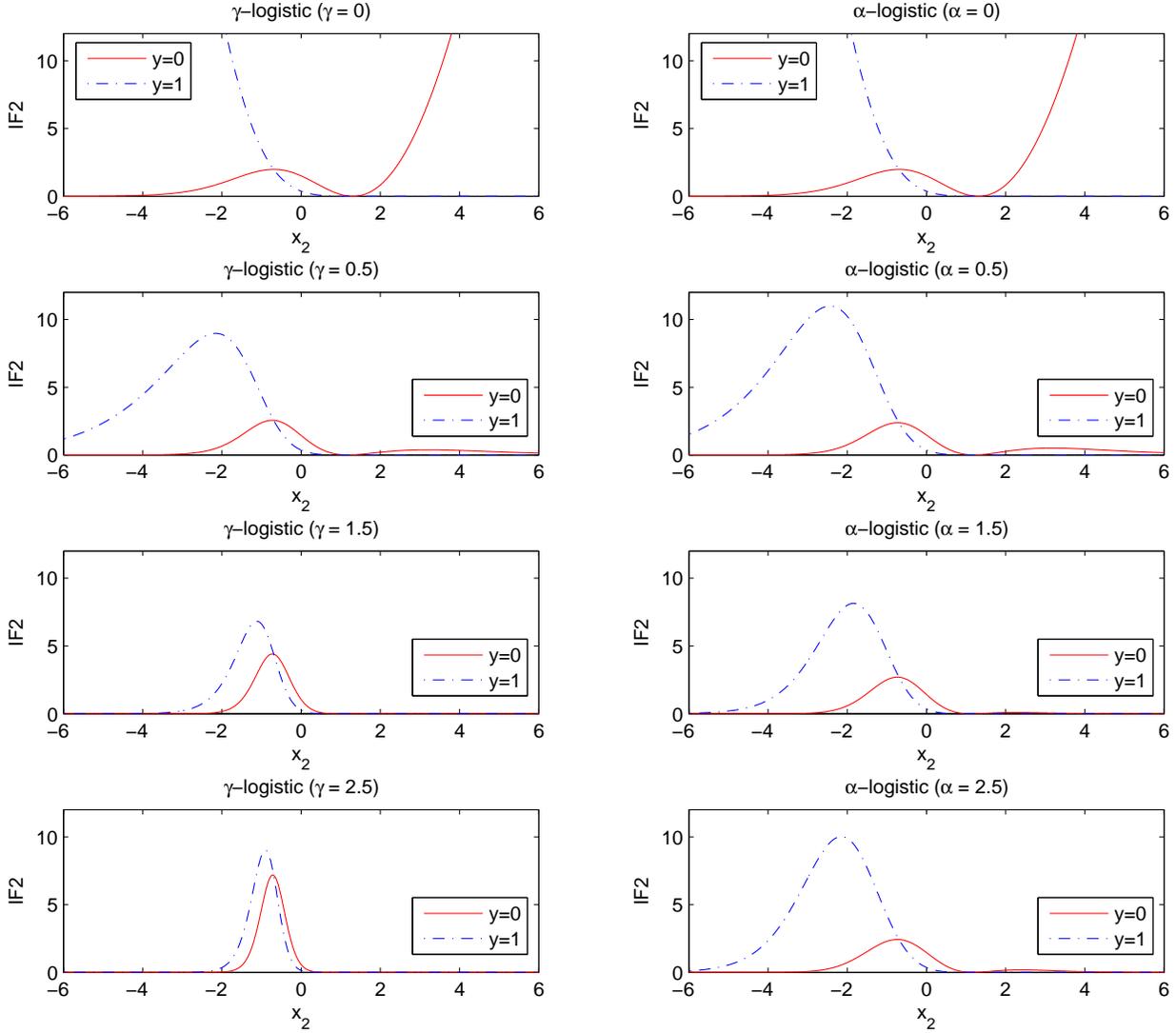}
\caption{The second-order influence functions of misclassification
rate of $\gamma$-logistic (the left panel) and $\alpha$-logistic
(the right panel) at $\gamma=\alpha\in\{0,0.5,1.5,2.5\}$, where the
real line is for the case of $y=0$, and the dash-dotted line is for
the case of $y=1$. The plots are obtained under the setting of
$p=2$, where $X_1=1$ is the intercept term, $X_2|Y_0=0\sim
N(-0.5,1)$, $X_2|Y_0=1\sim N(0.5,1)$ and $P(Y_0=1)=2P(Y_0=0)$. It
gives $\beta_0=(\ln2,1)^\top $. }\label{fig.IF2}
\end{figure}

The difference between two methods can be further clarified via
comparing the misclassification rate of the prediction rule
$y=I(\widehat\beta_\bullet^\top x>0)$, where $\widehat\beta_\bullet$
can stand for either $\widehat\beta_\gamma$ or
$\widehat\beta_\alpha$. Croux, Haesbroeck and Joossens (2008) showed
that the robustness of misclassification rate is characterized by
its second order influence function ${\sf
IF2}_{\widehat\beta_\bullet}(x,y)$. The second order influence
function for a functional $T(F)$ of the distribution $F$ at $z$ is
$\frac{\partial^2}{\partial\varepsilon^2}T\{(1-\varepsilon)F
+\varepsilon\delta_z\}|_{\varepsilon=0}$, where $\delta_z$ is the
Dirac measure at $z$. For the case of $p=2$ with $X_1=1$,
$\beta_0=(\beta_{01},\beta_{02})^\top $, and
$\widehat\beta_\bullet=(\widehat\beta_{\bullet
1},\widehat\beta_{\bullet 2})^\top $, one has ${\sf
IF2}_{\widehat\beta_\bullet}(x,y)\propto \{\beta_{01}\,{\sf
IF}_{\widehat\beta_{\bullet 2}}(x,y)-\beta_{02}\,{\sf
IF}_{\widehat\beta_{\bullet 1}}(x,y)\}^2$, which is plotted in
Figure~\ref{fig.IF2} with various $\gamma=\alpha$ values, where
${\sf IF}_{\widehat\beta_{\bullet j}}(x,y)$ is the influence
function of $\widehat\beta_{\bullet j}$, $j=1,2$. When $\gamma=0$,
both methods reduce to the non-robust MLE, and an unbounded ${\sf
IF2}_{\widehat\beta_\bullet}(x,y)$ is detected. Note that ${\sf
IF2}_{\widehat\beta_{\bullet}}(x,y)$ has larger value at non-matched
$(x,y)$ value, which reflects the influence of outliers. We also
detect that ${\sf IF2}_{\widehat\beta_{\bullet}}(x,0)>0$ around
$x=-1$. This is reasonable since $P(Y_0=1)=2P(Y_0=0)$ in our
setting, which gives more samples with $Y_0=1$. As a result, a data
point from the $0$-group is expected to be more influential than
that from the $1$-group. When $\gamma=0.5$, ${\sf
IF2}_{\widehat\beta_\bullet}(x,y)$ at non-matched $(x,y)$ are
largely reduced, indicating the robustness of $\gamma$-logistic and
$\alpha$-logistic to mislabeling. The difference between two methods
becomes clear when $\gamma\ge 1.5$, where ${\sf
IF2}_{\widehat\beta_\alpha}(x,y)>0$ for a wide range of $x$, while
${\sf IF2}_{\widehat\beta_\gamma}(x,y)>0$ at limited region of $x$
only. That is, $\gamma$-logistic becomes more and more resistant to
mislabeling as $\gamma$ increases. The robustness of
$\gamma$-logistic, as mentioned in Section~\ref{sec.div}, comes from
the locality nature of $\gamma$-divergence. It also implies that,
when $\gamma$ is large, the performance of $\gamma$-logistic is
mainly determined by data points near the decision boundary
$\beta_0^\top x=0$ ($x=-\ln2$ in this case). This explains the
observation at $\gamma\ge 1.5$ that ${\sf
IF2}_{\widehat\beta_\gamma}(x,y)$ can have larger value than ${\sf
IF2}_{\widehat\beta_\alpha}(x,y)$, especially when $y=0$ (i.e., the
$0$-group with fewer samples).

\begin{rmk}
There exist robustfied logistic regression methods other than the
constant-mislabel logistic, $\xi$-logistic, and $\alpha$-logistic. A
majority of them have a robust estimating equation of the form
$\frac{1}{n}\sum_{i=1}^n[w_i(\beta)\{Y_i-\pi(X_i;\beta)\}-b(X_i;\beta)]X_i=0$,
where the weight $w_i(\beta)$ can depend on $(X_i,Y_i)$. The bias
correction term $b(X_i;\beta)$ is used to ensure Fisher consistency
in the presence of $w_i(\beta)$. See Bianco and Yohai (1996),
Carroll and Pederson (1993), Stefanski, Carroll, and Ruppert (1986),
K\"{u}nsch, Stefanski, and Carroll (1989) among others for different
choices of $w_i(\beta)$. Note that $\gamma$-logistic does not belong
to this class, since it uses $\{Y_i-\pi(X_i;(\gamma+1)\beta)\}$ for
bias correction.
\end{rmk}

\section{Numerical Studies}\label{sec.sim}

\subsection{Simulation settings}\label{sec.sim1}

We use the Pima data (see Section~\ref{sec.data} for details) to
conduct simulation studies. In each simulation run, $n=500$
covariate vectors $X_0\in \mathbb{R}^8$ are randomly sampled from
the Pima data (after component-wise standardization) and
$X=(X_0^\top ,1)^\top $. Given $X$, the response variable $Y$ is
generated from (\ref{model.mis}) with the following settings of
mislabel probabilities: (S1) $\eta_0(x)=u_0$ and $\eta_1(x)=u_1$;
(S2) $\eta_0(x)=\eta_1(x)=u_0+(u_1-u_0)\, \frac{\exp(\beta_0^\top
x)}{1+\exp(\beta_0^\top x)}$; (S3)
$\eta_j(x)=u_0+(u_1-u_0)\frac{\exp(b_j^\top x)}{1+\exp(b_j^\top
x)}$, where each element of $b_j\in \mathbb{R}^{9}$, $j=0,1$, is
generated from $N(0,2^2)$ for each simulation; and (S4)
$\eta_0(x)=u_0+(u_1-u_0)I(|X_1-a|<3, |X_3+a|<3)$ and
$\eta_1(x)=u_0+(u_1-u_0)I(|X_1+a|<3, |X_2+a|<3)$, where $a\sim
N(2,0.3^2)$ for each simulation. Setting~(S1) considers
$Y_0$-dependent mislabeling. Setting~(S2) considers $X$-dependent
mislabeling, where mislabeling is more likely to occur for subjects
with higher success probability. Settings (S3)-(S4) consider
$(Y_0,X)$-dependent mislabeling. In (S3) $\eta_j(x)$'s depend on
random linear combinations of $X$. In (S4) mislabeling is more
likely to occur for $(X_1,X_3)$ around $(a,-a)$ when $Y_0=0$, and
also more likely to occur for $(X_1,X_2)$ around $(-a,-a)$ when
$Y_0=1$. We set $u_0=0.05$ and $u_1\ge 0.05$ such that in all
settings, $u_1=u_0$ indicates that the constant-mislabel logistic
holds, while $u_1>u_0$ indicates a deviation from it.

Two types of $\gamma$ selection are implemented. One is based on the
data-adaptive method in Remark~\ref{rmk.select_gamma} (denoted by
$\gamma$-logistic). The other is based on an independent
uncontaminated data $\{(X_i^*, Y_{0i}^*)\}_{i=1}^n$ that selects
$\gamma$ by maximizing the likelihood
$\prod_{i=1}^n\pi(X_i^*;\widehat\beta_\gamma)^{Y_{0i}^*}
\{1-\pi(X_i^*;\widehat\beta_\gamma)\}^{1-Y_{0i}^*}$ (denoted by
$\gamma^*$-logistic). Of course $Y_{0i}^*$'s are not observed, and
$\gamma^*$ only represents an ideal $\gamma$ value for comparison
purpose. In addition to $\gamma$-logistic and $\gamma^*$-logistic,
we also implement the conventional logistic regression (denoted as
logistic), constant-mislabel logistic, $\xi$-logistic, and
$\alpha$-logistic (where $\alpha$ is optimally tuned as
$\gamma^*$-logistic does, and it is denoted by $\alpha^*$-logistic).
Simulation results are reported with 500 replicates.

\subsection{Simulation results}\label{sec.sim2}

We first evaluate the performances of
$(\widehat\beta_\gamma,\widehat\Sigma_\gamma)$. Simulation results
for $\gamma=2$ under (S1)-(S2) with $\beta_0=(0,1,-1,1,
\0_{p-3}^\top )^\top $ and $u_1=0.1$ are placed in
Table~\ref{sim.tab}, which reports the means of
$\widehat\beta_\gamma$ (Mean), the standard deviations of
$\widehat\beta_\gamma$ (SD), and the means of the diagonal elements
of $\widehat\Sigma_\gamma$ (SE) over 500 replicates. One can see
that $\widehat\beta_\gamma$ targets $\beta_0$ with only small bias
under both mislabeling mechanisms (S1)-(S2). Moreover, SE are found
to be close to SD, which shows the validity of the proposed
sandwich-type estimator $\widehat\Sigma_\gamma$.

\begin{figure}[!ht]
\includegraphics[width=3.5in,height=2.8in]{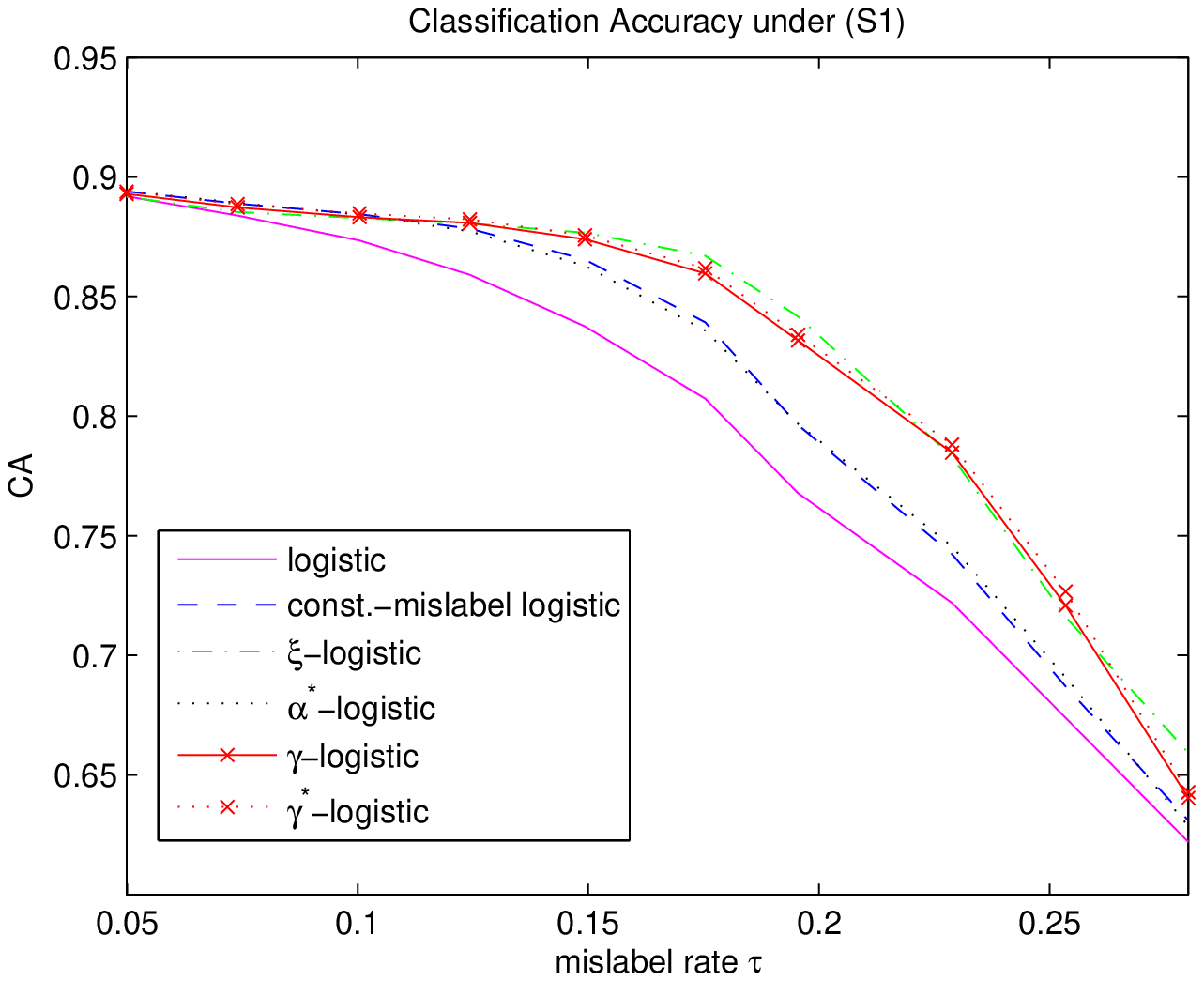}
\includegraphics[width=3.5in,height=2.8in]{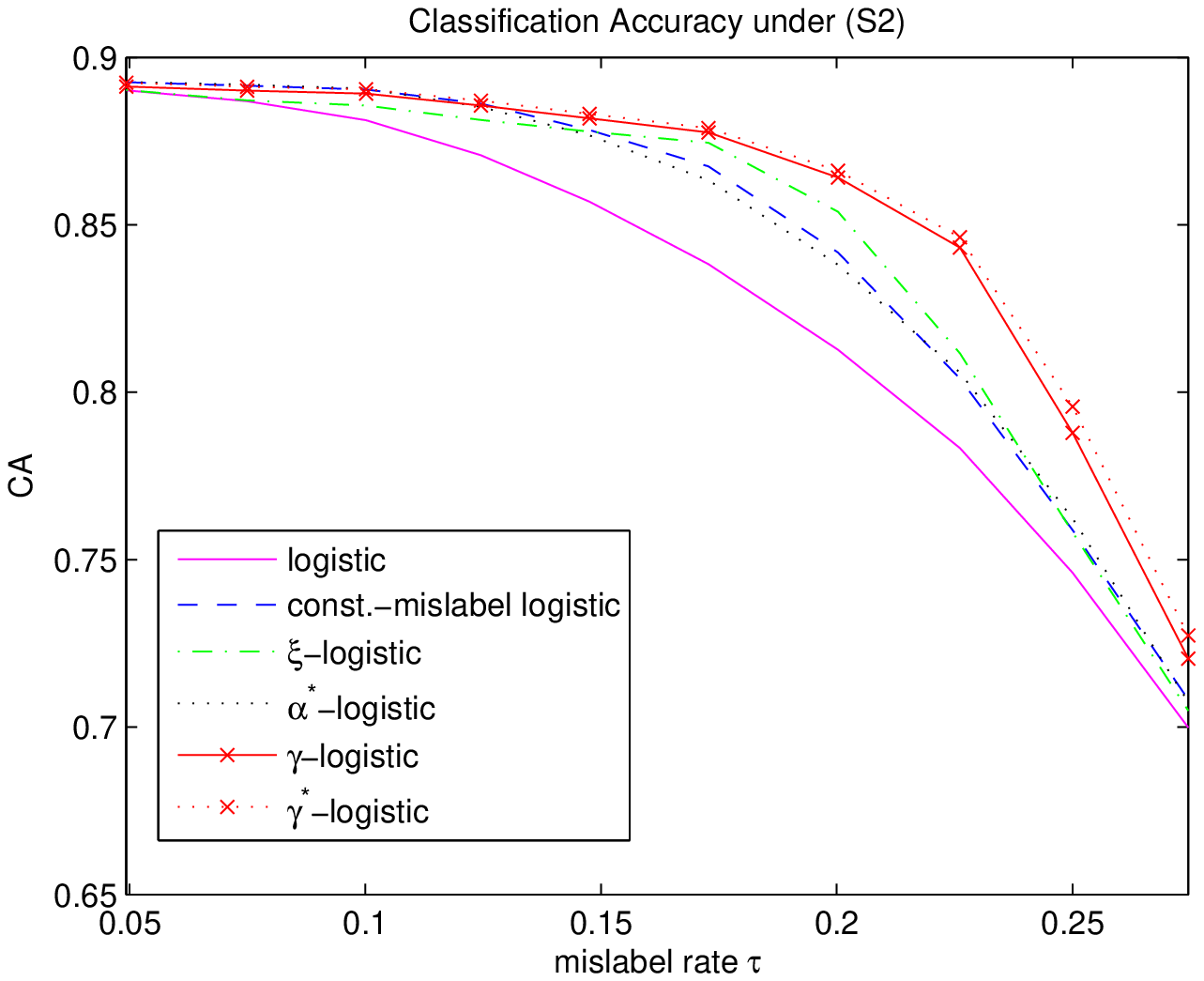}
\includegraphics[width=3.5in,height=2.8in]{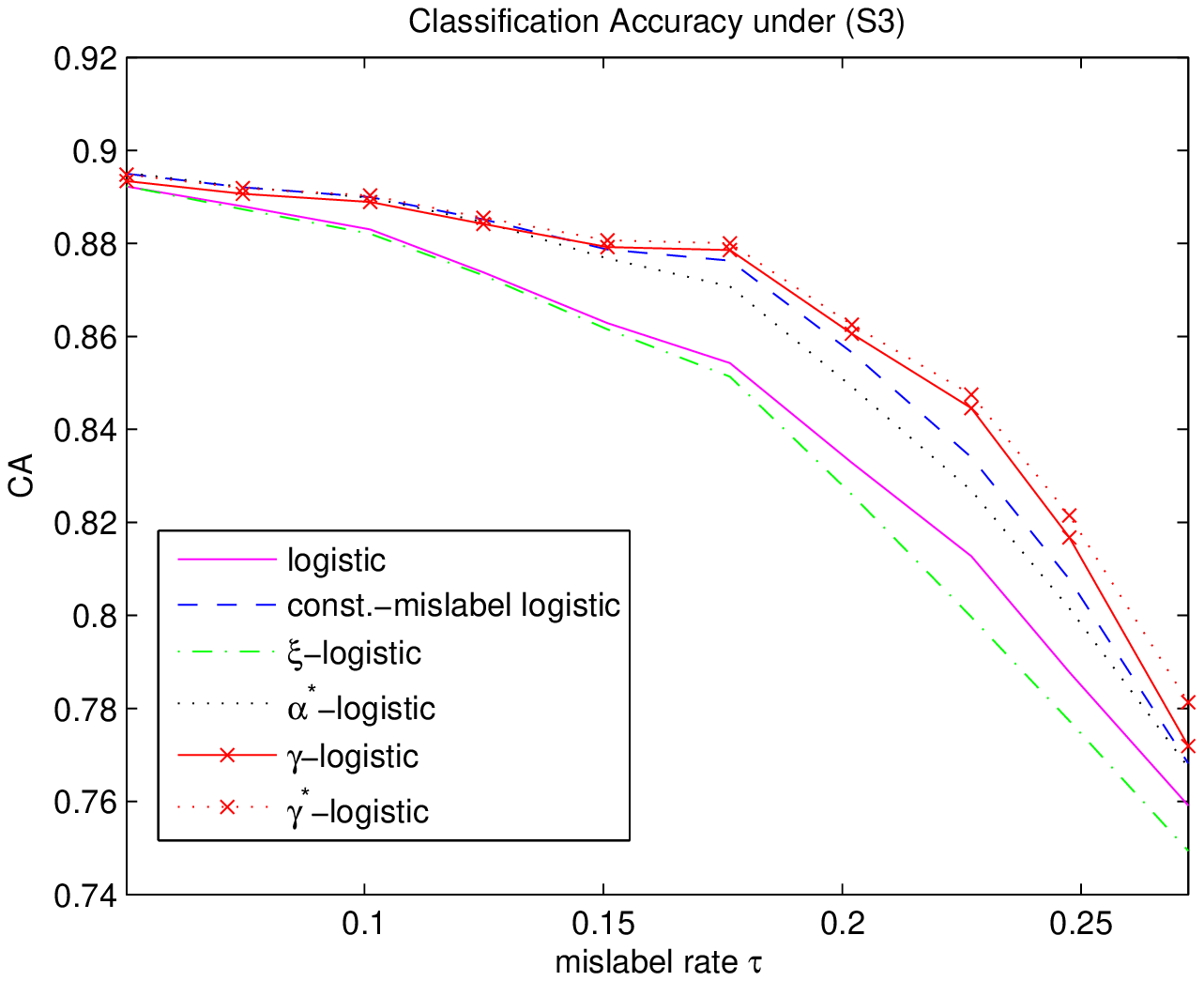}
\includegraphics[width=3.5in,height=2.8in]{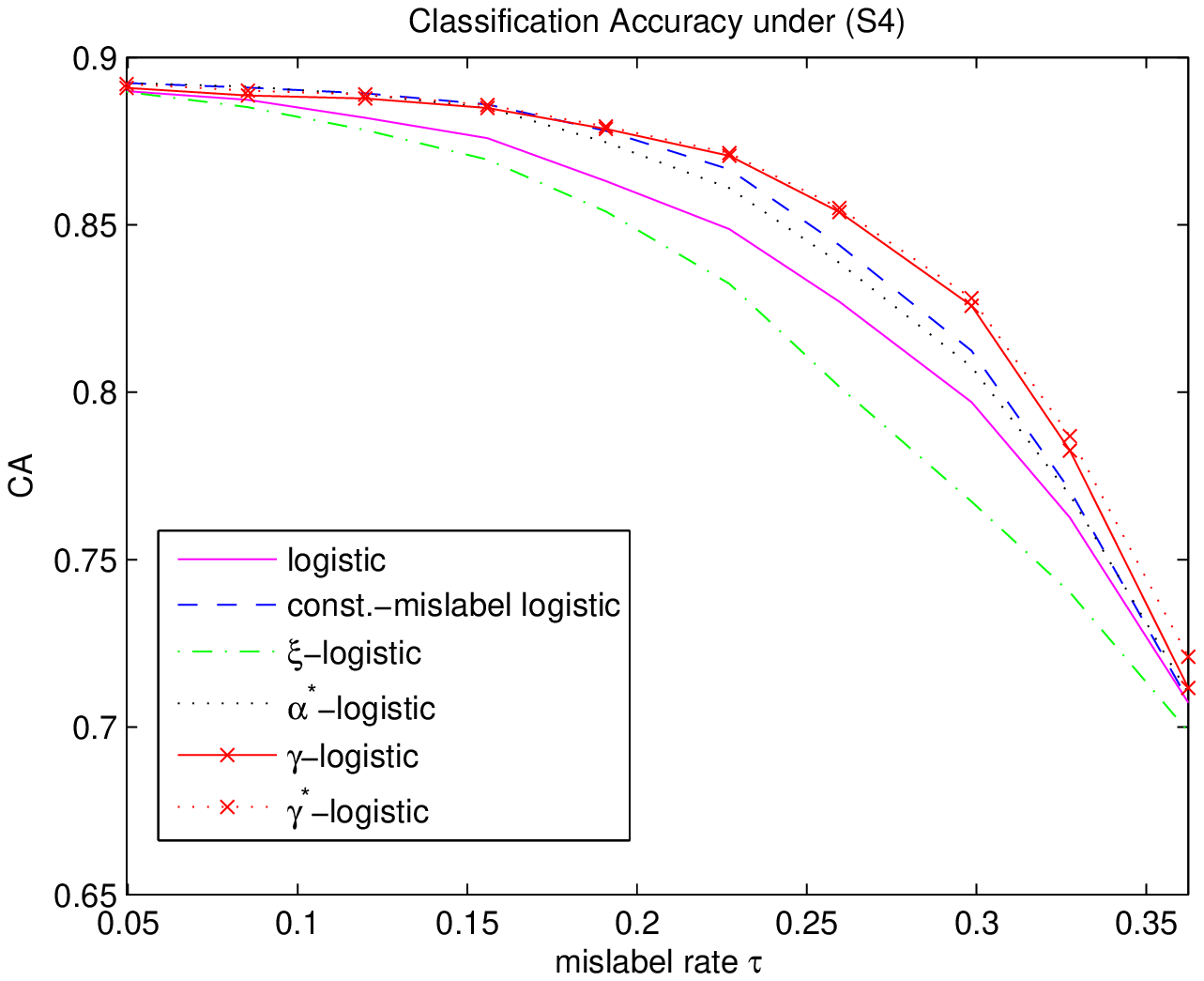}
\caption{Simulation results of the classification accuracy (CA)
under (S1)-(S4) with different values of $u_1$, where the $x$-axis
represents the corresponding mislabel rate $\tau=P(Y\ne
Y_0)$.}\label{fig.sim.1}
\end{figure}

We next compare the performance of $\gamma$-logistic with other
methods. The values of $\gamma$ and $\alpha$ are selected over
$[0.5,2.5]$ with $\gamma_0=0.1$. In this simulation, each element of
$\beta_0$ is generated from $N(0,2^2)$ for each replicate.
Figure~\ref{fig.sim.1} reports the classification accuracy (CA) from
applying the prediction rule $y=I(\widehat\beta_\gamma^\top x>0)$ to
an independent clean data $(Y_0,X)$ with size $n$, where the
$x$-axis represents the corresponding mislabel rate $\tau=P(Y\ne
Y_0)$ under $u_1\in\{0.05,0.1,\ldots,0.5\}$. We also report in
Table~\ref{sim.tab_gamma} the means of the selected $\gamma$ and
$\gamma^*$ values of $\gamma$-logistic and $\gamma^*$-logistic.
Observe that the robustified logistic methods ($\gamma$-logistic,
$\alpha$-logistic, constant-mislabel logistic) dominate the
conventional logistic under (S1)-(S4), but not the $\xi$-logistic.
Recall that $\xi$-logistic assumes that mislabeling tends to occur
for subjects lying near the decision boundary $\beta_0^\top x=0$.
This assumption is not satisfied in (S1)-(S4). As a result,
$\xi$-logistic can perform even worse than the conventional logistic
regression under (S3)-(S4), especially for the case of severe
mislabeling (i.e., large $\tau$). It conveys an important message
that, while incorporating a correct mislabeling mechanism into the
estimation method can be beneficial, the correctness of model
specifications for $\eta_j(x)$'s is critical to the analysis result.
Misspecifying $\eta_j(x)$'s can sometimes lead to worse result.
However, $\gamma$-logistic, which avoids modeling $\eta_j(x)$'s, is
able to adapt to various mislabeling mechanisms and can be less
affected by model misspecification.

We now compare $\gamma$-logistic with constant-mislabel logistic and
$\alpha$-logistic. For small $\tau$, the constant-mislabel
assumption approximately holds and constant-mislabel logistic
produces the highest CA values as expected, while $\gamma$-logistic
has comparable performances. For large $\tau$, the mislabeling
mechanism becomes complicated, which adversely affects the
performances of constant-mislabel logistic. In this case,
$\gamma$-logistic produces the highest CA values under (S1)-(S4). It
is also found that $\gamma$-logistic outperforms
$\alpha^*$-logistic, even $\alpha^*$-logistic selects $\alpha$
optimally. Recall the comparison discussions of robustness for
$\gamma$-logistic and $\alpha$-logistic in
Section~\ref{sec.comparison.alpha}. Our simulation results confirm
the superiority of $\gamma$-logistic in dealing with various
mislabeling mechanisms. Finally, comparing $\gamma$-logistic with
the optimal $\gamma^*$-logistic, the loss of $\gamma$-logistic from
using the data-adaptive $\gamma$ is not large, indicating the
applicability of the proposed data-adaptive selection criterion of
$\gamma$.

\section{The Pima Data Analysis}\label{sec.data}

The Pima data (available from the {\it UCI machine learning
repository}) contains females of Pima Indian heritage, each with an
indicator of diabetes status ($Y$) and 8 covariates (standardized to
have mean 0 and variance 1), including the pregnant times ($X_1$),
glucose concentration ($X_2$), blood pressure ($X_3$), triceps skin
fold thickness ($X_4$), serum insulin ($X_5$), BMI ($X_6$), diabetes
pedigree function ($X_7$), and age ($X_8$). We set $X_9=1$ to
include the intercept term. Detailed description of the data can be
found in Smith \textit{et al.} (1988). Medical data can more easily
suffer the problem of mislabeling, and we aim to use the robust
$\gamma$-logistic to investigate the effects of these covariates on
the diabetes status.

\begin{figure}[!ht]
\centering
\includegraphics[width=5in]{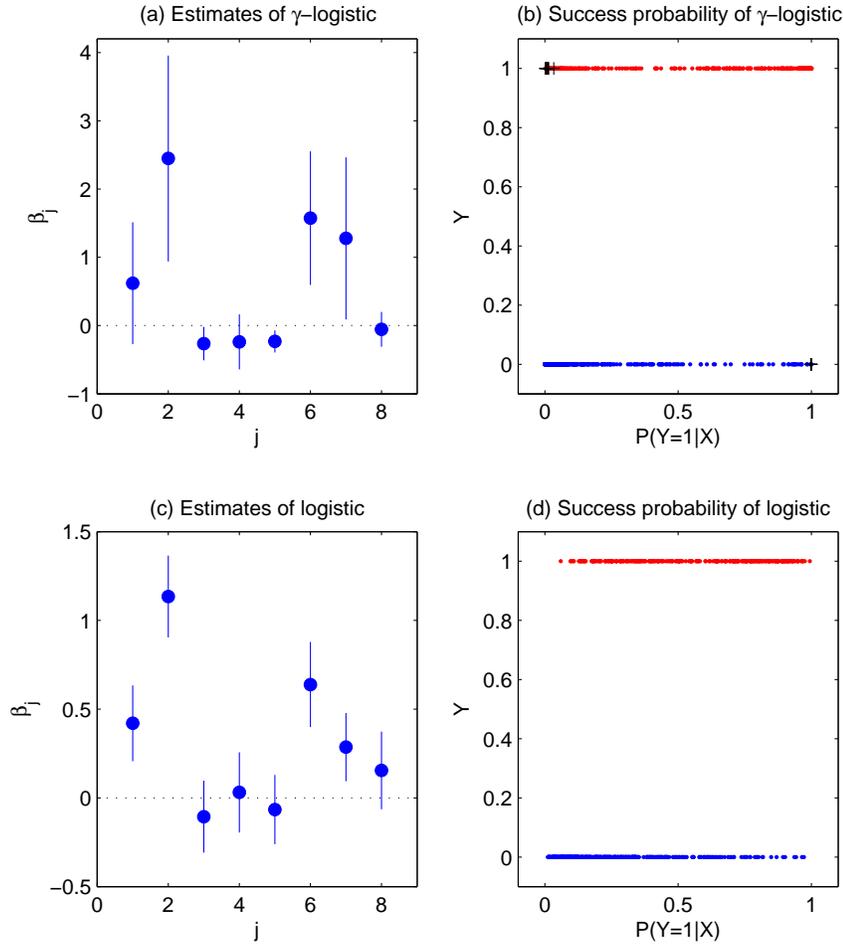}
\caption{(a): The regression coefficients $\widehat\beta_\gamma$
from $\gamma$-logistic, where the vertical lines represent the
$95\%$ confidence intervals. (b) The success probabilities
$\pi(X_i;\widehat\beta_\gamma)$'s for two groups $Y_i=1$ and $Y_i=0$
from $\gamma$-logistic. Subjects with $PV_i<0.01$ are marked with
``+''. The analysis results from conventional logistic are placed in
(c)-(d).}\label{fig.pima}
\end{figure}

Figure~\ref{fig.pima}~(a) provides the estimates
$\widehat\beta_\gamma$ from $\gamma$-logistic together with the
$95\%$ confidence intervals. Figure~\ref{fig.pima}~(b) provides the
estimated success probabilities $\pi(X_i;\widehat\beta_\gamma)$'s
for two groups. The analysis results from the conventional logistic
regressions are also placed in Figure~\ref{fig.pima} (c)-(d) for
comparison. In general, $\gamma$-logistic tends to produce wider
confidence intervals than conventional logistic. This is expected
since the robustness of $\gamma$-logistic comes at the cost of being
less efficient than MLE. Both analysis results show that
$(X_1,X_2,X_6,X_7)$ are critical (significant or nearly significant)
factors to the diabetes status. Interestingly, $\gamma$-logistic
further demonstrates that $(X_3,X_5)$ are significant factors (as
the corresponding confidence intervals do not contain 0), and $X_4$
is nearly significant. Considering the robustness of
$\gamma$-logistic, this difference would mainly result from treating
some instances as outliers, by assigning them less weights during
model fitting. In particular, we obtain more precise estimates for
the effects of blood pressure ($X_3$), triceps skin fold thickness
($X_4$), and serum insulin ($X_5$) when possible mislabeled subjects
have been weighed down.

From the results of $\gamma$-logistic, instances with $PV_i< 0.01$
are marked with ``$+$'' in Figure~\ref{fig.pima}~(b). These
instances are candidates of mislabeled subjects. To further
investigate the driven factors of mislabeling, we define the
mislabeling status $\delta_i=I(PV_i< 0.01)$, and then estimate the
true response by $\widehat Y_{0i}=Y_i(1-\delta_i)+(1-Y_i)\delta_i$,
i.e., subjects with $\delta_i=1$ are flipped for label correction.
We then fit the conventional logistic regression to
$(\delta_i,X_i)|\widehat Y_{0i}=j$ to obtain the regression
coefficient $b_j$ for $j=0,1$. Note that $b_j$ quantifies how $X$
affects the chance of being mislabeled within the group of $Y_0=j$.
The AUC values from $(\delta_i, b_0^\top X_i)|\widehat Y_{0i}=0$ is
0.713, while it is 0.925 from $(\delta_i, b_1^\top X_i)|\widehat
Y_{0i}=1$. It indicates that $X$ is influential to the mislabel
probability $\eta_1(x)$, while the mislabel probability $\eta_0(x)$
tends to be constant for subjects without diabetes. Moreover, the
result of $b_1=(0.258,0.322, 0.491, -0.837,0.303, 1.106, 1.061,
0.512, -6.989)$ suggests $(X_6,X_7)$ (with p-values smaller than
0.05) to be possible driven factors of mislabeling for diabetes
patients.

\section{Discussion}

In this work we only consider the case of mislabeling in the
response $Y$, while $X$ is assumed to be uncontaminated. In the
presence of leverage points of $X$ that are influential to the final
estimates, $\gamma$-logistic can be modified to mitigate the effects
of outlying $X_i$ by using a weighting scheme
$w_{\gamma,i}(\beta)q(X_i)$ in the estimating equation
(\ref{estimating_eq.gamma}). For example, Croux, Haesbroeck and
Joossens (2008) suggested $q(x)=I\{(x-\mu_X)^\top
\Sigma_X^{-1}(x-\mu_X)\le a\}$ for some user-defined constant $a$,
where $\mu_X$ and $\Sigma_X$ are some robust estimates of $E(X)$ and
$\cov(X)$. Since $q(X_i)$ does not depend on $Y_i$,
Theorem~\ref{thm.gamma} still holds for the modified
$\gamma$-logistic by replacing $H_\gamma$ and $U_\gamma$ with
$E[q(X_i)\|f(\cdot|X_i;\beta_0)\|_{\gamma+1}
\nu(X_i;(\gamma+1)\beta_0) X_iX_i^\top ]$ and $E[q^2(X_i)
w_{\gamma,i}^2(\beta_0)\{Y_i-\pi(X_i;(\gamma+1)\beta_0)\}^2
X_iX_i^\top ]$, respectively. It is of interest to investigate the
choice and effect of $q(\cdot)$ on $\gamma$-logistic in a future
study.

For the purpose of robustness, Ghosh and Basu (2016) developed a
robust GLM using the density power divergence, which includes
$\alpha$-logistic as a special case. We have shown that
$\gamma$-logistic outperforms $\alpha$-logistic under severe
mislabeling in numerical studies. The developed methodology
(\ref{model.pmf})-(\ref{gamma_div.reg}) can be extended to robust
GLM, including multi-class $Y$ (see Supplementary Materials for a
brief illustration), count $Y$, and continuous $Y$. Although the
idea is straightforward, further efforts are required to develop the
validity of the approximation (\ref{div.gamma.approx}), the
asymptotic properties, and the implementation algorithms. It is also
of interest to compare the differences between the robust GLM using
$\gamma$-divergence and the robust GLM of Ghosh and Basu (2016)
using density power divergence.

\section*{References}
\begin{description}

\item
Bianco, A. M. and Yohai, V. J. (1996). Robust estimation in the
logistic regression model. \textit{In Robust statistics, data
analysis, and computer intensive methods} (pp. 17-34). Springer
New York.




\item
Carroll, R. J. and Pederson, S. (1993). On robustness in the
logistic regression model. {\it Journal of the Royal Statistical
Society}, Series B, 55, 693-706.

\item
Copas, J. B. (1988). Binary regression models for contaminated
data. {\it Journal of the Royal Statistical Society}, Series B,
50, 225-265.


\item
Croux, C., Haesbroeck, G., and Joossens, K. (2008). Logistic
discrimination using robust estimators: an influence function
approach. {\it Can. J. Stat.}, 36, 157-174.



\item
Fujisawa, H. and Eguchi, S. (2008). Robust parameter estimation
with a small bias against heavy contamination. {\it Journal of
Multivariate Analysis}, 99, 2053-2081.

\item
Ghosh, A. and Basu, A. (2016). Robust estimation in generalized
linear models: the density power divergence approach. {\it
Test}, 25, 269-290.

\item
Hayashi, K. (2012). A boosting method with asymmetric
mislabeling probabilities which depend on covariates. {\it
Computational Statistics}, 27, 348-356.


\item
Jones, M. C., Hjort, N. L., Harris, I. R. and Basu, A. (2001). A
comparison of related density-based minimum divergence
estimators. {\it Biometrika}, 88, 865-873.

\item
Kanamori, T. and Fujisawa, H. (2015). Robust estimation under
heavy contamination using unnormalized models. {\it Biometrika},
102, 559-572.

\item
Komori, O., Eguchi, S., Ikeda, S., Okamura, H., S., Ichinokawa,
M., and Nakayama, S. (2016). An asymmetric logistic regression
model for ecological data. {\it Methods in Ecology and
Evolution}, 7, 249-260.

\item
K\"{u}nsch, H. R., Stefanski, L. A., and Carroll, R. J. (1989).
Conditionally unbiased bounded-influence estimation in general
regression models, with applications to generalized linear
models. {\it Journal of the American Statistical Association},
84, 460-466.

\item
Mollah, M. N. H., Eguchi, S., and Minami, M. (2007). Robust
prewhitening for ICA by minimizing $\beta$-divergence and its
application to FastICA, {\it Neural Process Lett.}, 25, 91-110.


\item
Stefanski, L. A., Carroll, R. J., and Ruppert, D. (1986).
Optimally bounded score functions for generalized linear models
with applications to logistic regression. {\it Biometrika}, 73,
413-424.

\item
Smith, J., Everhart, J., Dickson, W., Knowler, W., and Johannes,
R. (1988). Using the ADAP learning algorithm to forecast the
onset of diabetes mellitus. {\it Proceedings of the Symposium on
Computer Applications and Medical Care}, 9, 261-265.

\item
Takenouchi, T. and Eguchi, S. (2004). Robustifying AdaBoost by
adding the naive error rate. {\it Neural Computation}, 16, 767-787.

\item
Wainer, H., Bradlow, E. T., and Wang, X. (2007). {\it Testlet
Response Theory and Its Applications}. Cambridge University
Press, New York.
\end{description}


\clearpage

\newpage

\begin{table}
\centering \caption{The means of $\widehat\beta_\gamma$ (Mean), the
standard deviations of $\widehat\beta_\gamma$ (SD), and the means of
the diagonal elements of $\widehat\Sigma_\gamma$ (SE) from
$\gamma$-logistic under settings (S1)-(S2).}\label{sim.tab}
\vspace{3ex}
\begin{tabular}{rrrrrrrrrrrrrrrrrrrrrr}
\hline
\multicolumn{1}{c}{}&\multicolumn{5}{c}{(S1)} && \multicolumn{2}{c}{(S2)} \\
\cline{2-5} \cline{7-10}
\multicolumn{1}{c}{}  &\multicolumn{1}{c}{True}  &\multicolumn{1}{c}{Mean} &\multicolumn{1}{c}{SD} &\multicolumn{1}{c}{SE}  &&\multicolumn{1}{c}{True}  &\multicolumn{1}{c}{Mean} &\multicolumn{1}{c}{SD} &\multicolumn{1}{c}{SE}   \\
\cline{2-5} \cline{7-10}
$\beta_{01}$   &   0.000   &   -0.126  &   0.171   &   0.171   &   &   0.000   &   -0.014  &   0.168   &   0.169   \\
$\beta_{02}$   &   1.000   &   1.009   &   0.308   &   0.323   &   &   1.000   &   0.999   &   0.307   &   0.326   \\
$\beta_{03}$   &   -1.000  &   -0.999  &   0.312   &   0.316   &   &   -1.000  &   -0.995  &   0.296   &   0.315   \\
$\beta_{04}$   &   1.000   &   1.014   &   0.307   &   0.320   &   &   1.000   &   0.984   &   0.281   &   0.317   \\
$\beta_{05}$   &   0.000   &   -0.025  &   0.208   &   0.206   &   &   0.000   &   0.011   &   0.201   &   0.206   \\
$\beta_{06}$   &   0.000   &   -0.033  &   0.224   &   0.192   &   &   0.000   &   -0.006  &   0.217   &   0.197   \\
$\beta_{07}$   &   0.000   &   0.013   &   0.209   &   0.210   &   &   0.000   &   -0.001  &   0.223   &   0.216   \\
$\beta_{08}$   &   0.000   &   0.008   &   0.185   &   0.176   &   &   0.000   &   -0.018  &   0.190   &   0.184   \\
$\beta_{09}$   &   0.000   &   0.004   &   0.211   &   0.203   &   &   0.000   &   0.014   &   0.220   &   0.208   \\
\hline
\end{tabular}
\end{table}

\clearpage

\begin{table}
\centering \caption{The means of the selected $\gamma$ and
$\gamma^*$ values at different $u_1$ under
(S1)-(S4).}\label{sim.tab_gamma} \vspace{3ex}
\begin{tabular}{cccccccccccc}
\hline
$u_1$   &       &   0.05    &   0.10    &   0.15    &   0.20    &   0.25    &   0.30    &   0.35    &   0.40    &   0.45    &   0.50    \\
\hline
(S1)    &   $\gamma^*$  &   0.97    &   1.27    &   1.60    &   1.82    &   2.01    &   2.12    &   2.10    &   2.02    &   1.72    &   1.20    \\
    &   $\gamma$    &   1.26    &   1.60    &   1.84    &   2.04    &   2.20    &   2.32    &   2.30    &   2.21    &   1.92    &   1.76    \\
\hline
(S2)    &   $\gamma^*$  &   0.99    &   1.28    &   1.60    &   1.82    &   2.02    &   2.16    &   2.22    &   2.21    &   1.87    &   1.36    \\
    &   $\gamma$    &   1.30    &   1.59    &   1.87    &   2.05    &   2.20    &   2.30    &   2.37    &   2.28    &   1.95    &   1.64    \\
\hline
(S3)    &   $\gamma^*$  &   1.02    &   1.26    &   1.56    &   1.79    &   2.00    &   2.14    &   2.17    &   2.22    &   2.07    &   1.77    \\
    &   $\gamma$    &   1.35    &   1.57    &   1.84    &   2.08    &   2.20    &   2.31    &   2.37    &   2.40    &   2.38    &   2.34    \\
\hline
(S4)    &   $\gamma^*$  &   0.96    &   1.38    &   1.78    &   2.04    &   2.29    &   2.39    &   2.39    &   2.32    &   2.04    &   1.55    \\
    &   $\gamma$    &   1.29    &   1.74    &   2.03    &   2.24    &   2.39    &   2.46    &   2.45    &   2.35    &   2.11    &   1.87    \\
\hline
\end{tabular}
\end{table}

\end{document}